\documentclass[conference]{IEEEtran}
\IEEEoverridecommandlockouts
\usepackage{fancyhdr}
\usepackage{cite}
\usepackage{amsmath,amssymb,amsfonts}
\usepackage{algorithmic}
\usepackage{graphicx}
\usepackage{color}
\usepackage{textcomp}
\usepackage{xcolor}
\usepackage{pifont}   
\usepackage{makecell}
\usepackage{hyperref}
\usepackage{booktabs}
\usepackage{listings}
\usepackage{xcolor}
\def\BibTeX{{\rm B\kern-.05em{\sc i\kern-.025em b}\kern-.08em
    T\kern-.1667em\lower.7ex\hbox{E}\kern-.125emX}}

\definecolor{deepblue}{rgb}{0,0,0.5}   
\definecolor{deepred}{rgb}{0.6,0,0}
\definecolor{deepgreen}{rgb}{0,0.5,0}

\newcommand{\cmark}{\ding{51}} 
\newcommand{\xmark}{\ding{55}}

\begin{document}

\title{A RISC-V Simulator for the Tenstorrent Tensix Architecture}

\author{\IEEEauthorblockN{Nick Brown}
\IEEEauthorblockA{\textit{EPCC,} \\
\textit{University of Edinburgh}\\
Edinburgh, UK \\
0000-0003-2925-7275}
\and
\IEEEauthorblockN{Daniyal Arshad}
\IEEEauthorblockA{\textit{EPCC,} \\
\textit{University of Edinburgh}\\
Edinburgh, UK \\
0009-0008-6523-7582}
}

\maketitle

\begin{abstract}
RISC-V based AI accelerators have become popular and not only for AI and ML, but also as vehicles for exploring HPC workloads more generally. Tenstorrent are one example, and whilst they document their Tensix architecture in detail and develop their software stack openly, running code has historically required a physical card, and their recently released simulator targets hardware equivalence rather than introspection. In this paper we present Wolfpine, a community developed RISC-V simulator for the Tensix architecture, supporting both the Wormhole and Blackhole generations and integrating seamlessly with the TT-Metal software stack. 

Constructed entirely from publicly available documentation, open source software and measurements on physical hardware, Wolfpine is written in pure Python for readability and ease of modification. An opt-in, provenance-tracked cost model adds timing fidelity incrementally, and typed event streams, per-component tracing and timeline export enable programmers to identify and diagnose performance bottlenecks that the hardware itself hides.
\end{abstract}


\section{Introduction}
RISC-V is a community driven open Instruction Set Architecture (ISA) and comprises a wide range of participants who are working to further develop extensions to the standard to tackle a wide range of workloads. At the same time, from CPUs to accelerators, we are seeing numerous vendors announcing and making available their RISC-V based products to the community. The High Performance Computing (HPC) community are keen to experiment with developing codes for these architectures \cite{brown2025risc}, yet do not always have access to the physical hardware all the time, and furthermore need detailed diagnostics for performance tuning which is not always exposed by the hardware.

Tenstorrent are an example of this with their Tensix-based accelerators which, whilst targetted mainly for AI workloads, have demonstrated promise in accelerating HPC codes more widely \cite{brown2024accelerating} \cite{almerol2025accelerating} and delivering impressive energy efficiency properties \cite{arshad2026spectral}. However, up until recently Tenstorrent did not have an open source simulator and whilst their hardware is widely available and at a reasonable price point, this limited development of codes \emph{offline} as well as people being able to experiment before committing to purchase. A potential solution however is that Tenstorrent have adopted an open way of working, releasing detailed documentation information about their architecture \cite{ttisadoc} and open sourced software stack \cite{ttmetal}. 

In this paper we describe Wolfpine which is an open source, community developed, simulator aimed at AI accelerators and currently implements the Tenstorrent Tensix architecture for both Blackhole and Wormhole. Developed entirely based on the open source documentation, software stack and measurements made on physical hardware, this provides not only an independent reference simulator but furthermore a wider range of functionality than Tenstorrent's own simulator for profiling, introspection and debugging codes. Moreover, given the open nature of RISC-V we have designed this as an extensible, modular, tool which can be extended in future to support additional architectures and RISC-V extensions.

\section{Background and related work}

Tenstorrent's accelerators are built around a two-dimensional grid of \emph{tiles}. This paper is focussed on two generations, the \emph{Wormhole}, whose grid contains up to 80 \emph{Tensix} compute tiles, and its successor the \emph{Blackhole}, which scales to 140 alongside larger memories and wider interfaces. Parts are commonly \emph{harvested}, with some tiles disabled for yield, so a given accelerator exposes fewer than its architectural maximum and the Blackhole used in this paper reports 120. The grid also contains DRAM tiles which provide off-chip GDDR6 memory channels (six channels totalling 12\,GiB on Wormhole, eight channels totalling approximately 32\,GiB on Blackhole), ethernet tiles for chip-to-chip communication, and tiles hosting the PCIe interface and management (ARC) processor. Communication between all these is undertaken via two independent Networks-on-Chip (NoCs), each of which is a two-dimensional unidirectional wrapping torus, with the two routing in opposite directions. Each tile attaches to each NoC through a NoC Interface Unit (NIU) and a single NoC transaction may read or write the memory of any tile on the device.

\begin{figure}[htb]
\centering
 \includegraphics[width=\columnwidth]{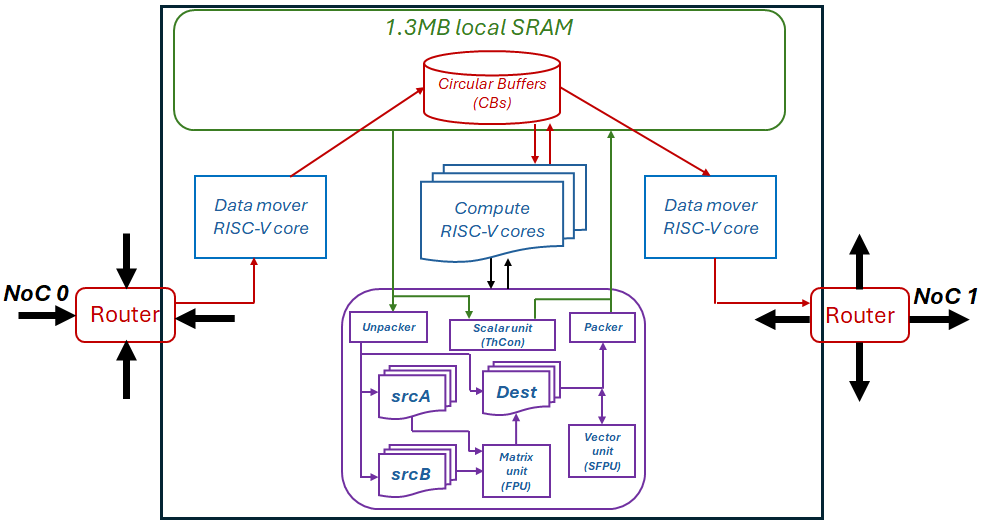}
\caption{Sketch of a single Tenxix block from \cite{brown2025exploring}}	
\label{fig:tensix}
\end{figure}
 
Figure \ref{fig:tensix} illustrates a Tensix compute tile in the Wormhole, each containing scratchpad SRAM known as \emph{L1} (1464\,KiB on Wormhole, 1536\,KiB on Blackhole), five small in-order 32-bit RISC-V cores referred to as \emph{baby cores}, and the \emph{Tensix co-processor}. The five RISC-V baby cores have fixed roles, two provide data movement (BRISC and NCRISC, one per NoC) and three called TRISC drive the Tensix co-processor. The co-processor itself provides the compute horsepower and is organised as a front end of three \emph{threads}, one paired with each of the TRISC cores which can issue instructions to the co-processor which is then dispatched to one of the \emph{back-end units}, potentially stalled, replayed multiple times, or expanded to a larger instruction stream in the frontend depending on the state of the co-processor. The co-processor backend units comprise the unpacker which moves and converts operands from L1 into the internal source registers SrcA and SrcB, the matrix unit (FPU) which performs matrix multiply and elementwise operations into the destination register file \emph{Dst}, the vector unit (SFPU) which operates lanewise on the Dst register, the packer which converts and writes results from Dst back to L1, a scalar/configuration engine (THCON), and a synchronisation unit controlling semaphores and thread interlocks. Numerous parts of the Tesix co-processor are exposed to the baby cores as memory-mapped ranges alongside the L1 memory itself, so that programming can be done entirely through loads and stores. Furthermore, there is a Tenstorrent bespoke RISC-V \emph{.ttinsn} instruction extension that issues instructions from TRISC to the Tensix co-processor.
 
Programming is undertaken through \emph{TT-Metal} (TT-Metalium) \cite{ttmetal}, Tenstorrent's open source software stack. A host program written against the TT-Metal API allocates buffers in DRAM or L1, and launches \emph{programs} that comprise kernels. There are typically three types of kernel; two data-movement kernels that execute on the data-movement cores (typically one for reading and one for writing), and a compute kernel which TT-Metal's toolchain splits across the three TRISCs (unpack, math and pack stages) to drive the co-processor cooperatively. Kernels communicate within a Tensix tile through \emph{circular buffers} which combine data storage and movement with synchronisation. 

\cite{cawleywormhole} was the first to document this architecture in detail, with Tenstorrent then publishing open ISA documentation ~\cite{ttisadoc}, describing the Tensix instruction set, the tiles' memory maps and the behaviour of the units summarised above. This documentation, together with the open TT-Metal and driver source code, constitutes the information from which the simulator described in this paper was developed.
 
\subsection{Related work}
\label{sec:related}
 
The most directly related system to our work is \texttt{ttsim}, Tenstorrent's own simulator~\cite{ttsim} that was released after we started this simulator development. \texttt{ttsim} provides a virtual Wormhole, Blackhole or (pre-silicon) Quasar device as a compiled shared library which TT-Metal loads in place of silicon, and positions itself as the golden reference implementation of the Tenstorrent architecture. It targets bit-exact numerical equivalence with silicon and is itself available as open source. This is a mature and broad in coverage simulator, running substantial portions of the TT-Metal, TT-NN and LLK test suites, with experimental multi-chip configurations. 

However, it is important to highlight that the work described in this paper is not in competition with \texttt{ttsim}, and we in fact use it for validation. Instead, our work differs in provenance, purpose and construction where \texttt{ttsim} has been developed internally by the vendor, with the public repository containing filtered source releases and accepting no external contributions. It is therefore authoritative, but is neither a codebase an outsider could have produced nor a repository setup for third-party modification. Our simulator demonstrates a complementary approach, exploring whether the published documentation and measurements on commodity hardware are sufficient to construct a working whole-device simulator. This helps test just how open and complete the vendor's documentation and source code actually is. Secondly, our simulator and \texttt{ttsim} make opposite engineering trade-offs where \texttt{ttsim} is C++ and compiled for speed, by contrast our simulator is pure Python and organised around the block abstraction of Section~\ref{sec:design}. Ultimately, we prioritise the legibility of design and code, ease of modification and experimentation, for example where adding new functional back-end units is a small overall change in our approach. Thirdly, \texttt{ttsim} concentrates on functional correctness and is limited in the information it provides about performance to the programmer. By contrast, the ability to identify and diagnose potential performance bottlenecks, which might be hidden from the programmer on the hardware itself, is a major aim of our simulator and as described in Section \ref{sec:ours} we provide an optional provenance-tracked cycle cost model over different parts of the architecture (Section~\ref{sec:cycle_accurate}). Moreover, observability is a primary concern and provided through typed event streams, per-component tracing, timeline and profile export (Section~\ref{sec:observability}).
 
More broadly, architectural simulation is a rich ecosystem used extensively in the RISC-V community. General-purpose full-system simulators, pre-eminently gem5~\cite{gem5}, provide modular, configurable models spanning functional to cycle-approximate detail, and have been extended toward accelerators through co-simulation frameworks such as gem5-Aladdin~\cite{gem5aladdin}. The Structural Simulation Toolkit~\cite{sst} also composes component models with an emphasis on scalable parallel simulation. These frameworks excel at design-space exploration enabling hardware designers to explore different customisations, such as the size of the cache. The goal of our work is fundamentally different, where instead we focus on providing fidelity to specific hardware, the Wormhole and Blachole, and this is closer in spirit to GPU simulators such as GPGPU-Sim and Accel-Sim~\cite{accelsim}. Those simulators model Nvidia GPUs faithfully enough to run CUDA binaries, and encounter a similar challenge to us where they must identify microarchitectural details about a commercial device, and in their case they in part leverage microbenchmarks. 

At the other ends of the fidelity spectrum sit analytical models of Deep Neural Network (DNN) accelerators such as Timeloop~\cite{timeloop} and SCALE-Sim~\cite{scalesim}, which estimate performance and energy from loop-nest and dataflow descriptions without executing software. RTL-level approaches, such as Gemmini~\cite{gemmini} and FPGA-accelerated simulation through FireSim~\cite{firesim}, offer cycle exactness for designs whose RTL is available, however this is often not the case for commercial accelerators. Functional ISA simulators such as Spike~\cite{spike} and QEMU~\cite{qemu} sit at the purely functional end, indeed our per-core commit logs are deliberately Spike-compatible to enable differential debugging of the RISC-V baby cores. Within this landscape, our simulator occupies a position we believe is scarcely populated, providing a whole-device, software-stack-faithful simulator of a shipping commercial AI accelerator, that has been constructed and calibrated exclusively from open sources, with timing fidelity and provenance made explicit.

\section{The Wolfpine architecture simulator}
\label{sec:ours}
\label{sec:design}
\begin{figure}[htb]
\centering
 \includegraphics[width=\columnwidth]{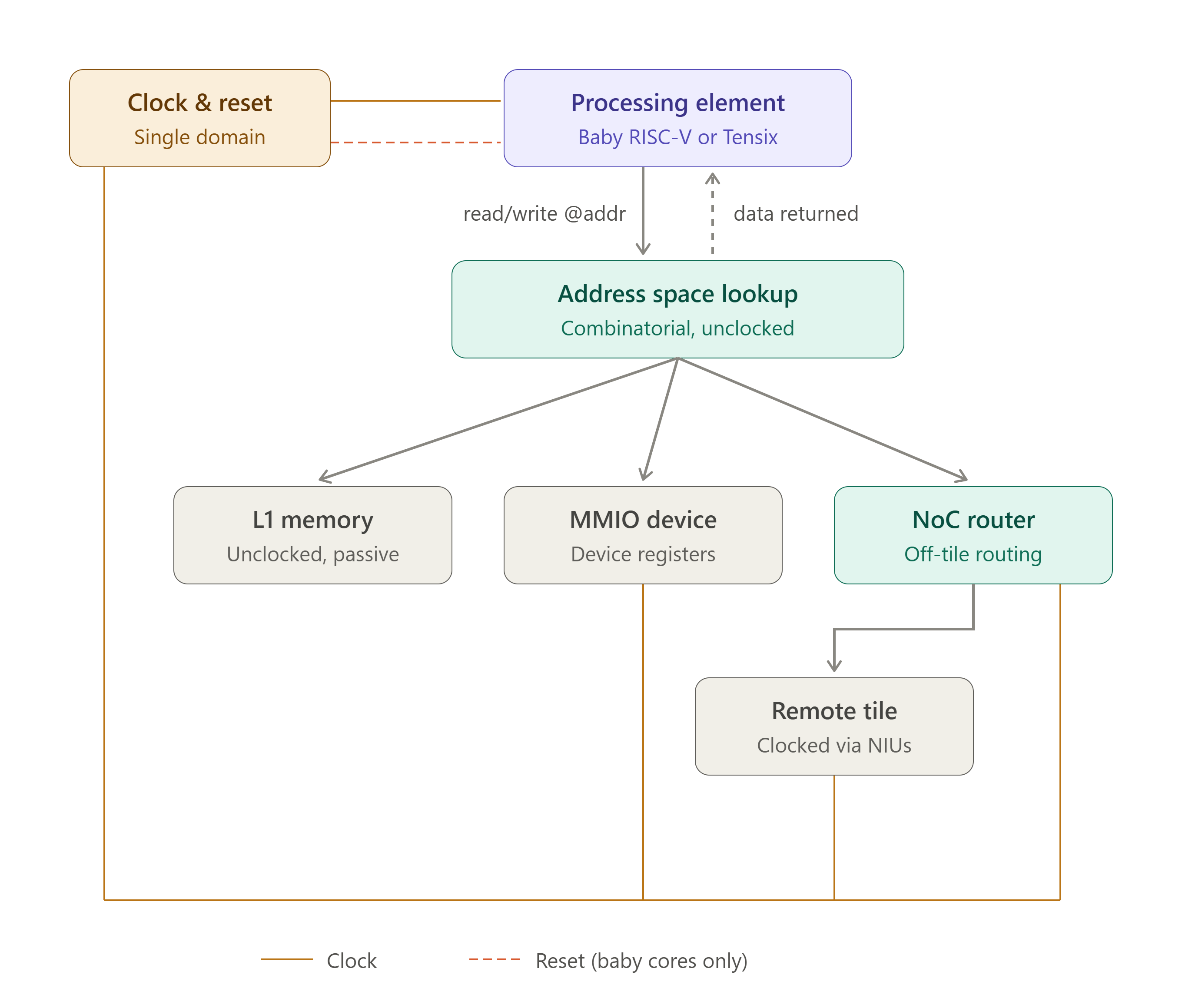}
\caption{Sketch of Wolfpine's conceptual architectural approach}	
\label{fig:general_approach}
\end{figure}

Conceptually, we base our abstraction upon the IP block design approach popularised by tools such as AMD Xilinx's Vivado. As sketched in Figure \ref{fig:general_approach}, Wolfpine's architecture is organised as a set of blocks, each of which is either active or passive. Active blocks are driven by a clock and, optionally, a reset. They implement specific methods which are invoked by the clock on every cycle, and by the reset when it changes state. Processing Elements (PEs) are a common type of active block, and examples include CPUs or the Tensix co-processor, which undertake some processing of instructions and/or data. The abstract class that must be implemented for the clock and reset are reported in Listings \ref{lst:clockable} and \ref{lst:resetable} respectively. Passive blocks, by contrast, implement no per-cycle behaviour and are actioned only on demand. Memory is an example of this where state is maintained but no work is undertaken of its own. Blocks are composed into tiles, and tiles into a device, with all the wiring which comprises the registration of address ranges, and membership of the clock and reset performed at construction time. This is effectively the programmatic equivalent of assembling a block design on a canvas in tools such as Xilinx's Vivado.

\begin{lstlisting}[language=Python, caption={The clock interface implemented by active blocks.}, label={lst:clockable}]
class Clockable(ABC):
    @abstractmethod
    def clock_tick(self, cycle_num):
        # Called by the clock on every cycle

    def is_clock_idle(self):
        # True when block is idle so clock
        # tick has no effect
        return False

    def next_wake_cycle(self, cycle_num):
        # Cycle at which this block next 
        # requires attention, None if 
        # only external stimulus change state
\end{lstlisting}

\begin{lstlisting}[language=Python, caption={The reset interface, implemented by active blocks which additionally participate in reset.}, label={lst:resetable}]
class Resetable(ABC):
    @abstractmethod
    def reset(self):
        # Called when the reset changes state
\end{lstlisting}

Ranges of memory addresses, each defined by a base address and a size, are registered for every block to direct memory accesses to their target. All accesses issued by a processing element pass through a lookup arbitrator which locates the block whose range covers the address, rebases the address relative to the start of that range, and hands the access to the target block to be actioned. Address spaces are layered rather than global, where each processing element is provided with its own view, containing ranges private to it, which is combined with the ranges shared across the tile. This is how, for instance, the private local memories of each baby RISC-V core can appear at an identical architectural address in every core yet resolve to distinct backing storage, whilst L1, the NoC Interface Unit (NIU) and the Tensix co-processor configuration space are shared across all. Ranges within a view are verified to be non-overlapping, and consequently lookup is both unambiguous and cheap.

From the perspective of the issuing processing element, a memory access completes synchronously where the target actions the read or write when it is handed the access, and any data is returned immediately, reflecting the functional rather than timing-driven nature of the simulator. As can be seen in the abstract class of Listing \ref{lst:memmapable}, which each memory element must implement, the important exception is that a target may instead respond with a stall, in which case the issuing core holds the current instruction and reattempts the access on subsequent cycles. This is the one point at which the passive memory path and the clocked world deliberately couple, and it is the mechanism by which blocking constructs such as mailboxes and synchronisation primitives are expressed without needing to model a bus protocol.

\begin{lstlisting}[language=Python, caption={The memory access interface implemented by any block registered in an address space. Addresses arrive rebased relative to the block's range; a read or write may return \texttt{MemoryStall}, causing the issuing core to hold and reattempt the access on subsequent cycles.}, label={lst:memmapable}]
class MemMapable(ABC):
    @abstractmethod
    def getSize(self):
        # Size in bytes of the 
        # addressable range

    @abstractmethod
    def read(self, addr, size):
        # Read operation, returning the data
        # or MemoryStall

    @abstractmethod
    def write(self, addr, value, size):
        # Write operation, may 
        # return MemoryStall
\end{lstlisting}

A consequence of this design, which we have found especially valuable, is its uniformity across the NoC. The NIU is simply another block registered in the tile's address space, and so remoteness is invisible to the processing element. Consequently, an off-tile access matches the NIU's address range, is carried across the NoC, and upon arrival at the destination tile re-enters exactly the same lookup-and-dispatch arbitration against that tile's own address space. The dispatch mechanism is therefore applied recursively rather than being a special case, and a DRAM read issued by a baby RISC-V core is the same mechanism applied twice, once locally to reach the NIU and once remotely to reach the memory.

Finally, a cycle consists of a deterministic, ordered invocation of the clock method of every registered active block, yielding an execution model that is cycle-driven but not cycle-accurate. State advances at the granularity of cycles, and ordering between components is well defined and reproducible, but the number of cycles consumed does not model hardware timing. Reset is modelled as a change of state rather than a pulse, mirroring the soft-reset register on the hardware. Notably, reset reaches only specific active blocks and passive blocks holding state survive it. This matches the persistence of L1 and DRAM contents across a soft reset on the physical device.

\begin{lstlisting}[numbers=none, language=Python, caption={Composing a minimal tile comprising a Wormhole baby RISC-V core, the Tensix co-processor and L1 connected through address range registration.}, label={lst:composition}]
# L1 SRAM
l1 = SRAM(L1_SIZE)

# The Tensix co-processor exposes memory
# mapped blocks for its interfaces
tensix = TensixCoProcessor()

mem_map = MemoryMap()
mem_map[AddressRange(0x0, l1.getSize())] = l1
mem_map[AddressRange(0xFFE40000, ...)] = \
    tensix.getThread(0)  # instruction push
tile_mem = TensixMemory(mem_map)

# RV babycore issues all accesses
trisc0 = WHBabyRISCV(BabyRISCVCoreType.TRISC0,
    [tile_mem], start_address=FW_BASE)

# Clock and reset drive active blocks only
clock = Clock(tensix.getClocks() + [trisc0])
reset = Reset([trisc0])

reset.reset()
clock.run(NUM_CYCLES)
\end{lstlisting}

Listing \ref{lst:composition} sketches top level integration of these different blocks, first creating local SRAM, before creating the tensix co-processor block by instantiating the \emph{TensixCoProcessor} class. The memory map, \emph{MemoryMap} is then created, placing the SRAM block from address \emph{0x0} and the tensix co-processor instruction push from \emph{0xFFE40000} as per \cite{ttisadoc}. The \emph{TensixMemory} class wraps one or more memory maps and undertakes decoding of addresses to locate target blocks across these maps and ranges. A Wormhole baby RISC-V core, the \emph{WHBabyRISCV}, is then instantiated, with the core type (in this case being the first of three cores to drive the tensix co-processor), the tile memory and its start address. Lastly, the clock and reset are created with appropriate blocks provided to them and the clock is then run for \emph{NUM\_CYCLES} number of cycles. Whilst this example provides a very simple sketch, it can be seen how these blocks and their interconnection through the memory system forms the central conceptual design of our simulator.

\subsection{Integration with TT-Metal}

\begin{figure}[htb]
\centering
 \includegraphics[width=0.75\columnwidth]{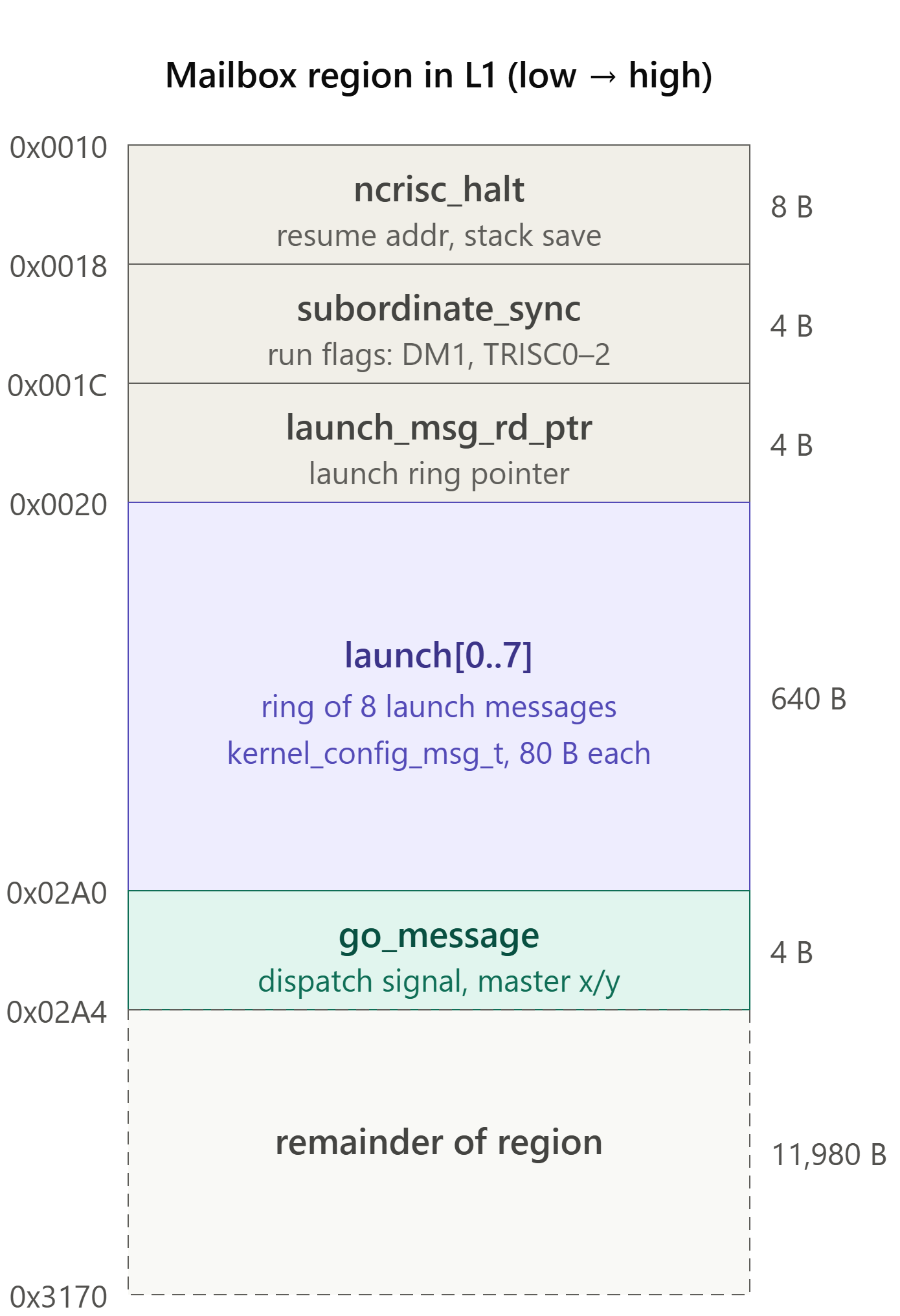}
\caption{Sketch of mailbox memory map through which the host runtime and firmware coordinate kernel execution}	
\label{fig:mem_map}
\end{figure}

Before a kernel is run on the Tenstorrent hardware, TT-Metal firmware is launched on each Tensix tile to configure the system. There are five firmware binaries, one for each baby RISC-V core within a tile, and the firmware sets up data structures which enables the host and device to meaningfully communicate, initialises different parts of the Tensix tile and then lastly enters a busy-waiting loop to spin for commands from the host. The mailbox memory map is sketched in Figure \ref{fig:mem_map} as it appears within each Tensix tile's L1. The region occupies 12,640 bytes beginning at \texttt{MEM\_MAILBOX\_BASE} (address \texttt{0x10}). Working from low to high addresses this comprises a halt record through which NCRISC saves and restores its resume address and stack when parked, a set of per-core synchronisation flags by which subordinate cores report their run state to BRISC, a read pointer and an eight-entry list of launch messages that each describe a kernel execution. Each of these includes runtime-argument offsets, circular-buffer configuration, NoC assignment and enable flags provided by the host and consumed by the device. Finally, the four-byte go message is held and it is this which the firmware waits upon and triggers the cores into executing the most recently written launch entry. None of this layout is hard-coded and the offsets, sizes and memory locations are determined by TT-Metal, with layouts changing between TT-Metal releases.

Initially integration with TT-Metal was done offline, where the mailbox configuration was written out into a YAML file and then Python driver code set this up and recreated central TT-Metal functionality to launch and manage kernel binaries across the simulated Tensix tiles. However, this was error prone and from the programmer's perspective rather clunky. Instead, it was desirable for TT-Metal and Wolfpine to integrate seamlessly, and there was already a route to achieving this based on Tenstorrent's own simulator which was released after our initial development. 

TT-Metal's user-mode driver provides a simulation backend which, when directed to do so via an environment variable, launches a simulator in place of physical silicon and communicates with it over a socket. We reverse engineered this wire protocol from the openly available driver source and found that it comprised a small command set carried as FlatBuffer-encoded messages over an \emph{nng\_pair} socket. Commands include opening and closing the device, performing memory reads and writes, and core reset control. We implemented the device side of this protocol in our simulator, and consequently an ordinary TT-Metal program now runs against it without any code changes, exactly as it would run on hardware. This also meant our simulator worked seamlessly with upstream TT-Metal, removing the need for a bespoke fork.

This transparency also provides a further development benefit. Because upstream TT-Metal example programs, in \emph{programming-examples}, validate their own results, they double as an end-to-end functional test suite for our simulator. This enables these existing codes to exercise the full path from host runtime through kernel dispatch, firmware, NoC transfers and Tensix computation. A set of these programs is maintained in the repository, built against an unmodified TT-Metal checkout, and run as part of routine testing.

\subsection{Fidelity with open source}
A central constraint we placed upon ourselves, and one which we believe makes this work repeatable for other architectures, is that the simulator is derived entirely from publicly available information. This combines the vendor's open ISA documentation, their open source TT-Metal software ecosystem, and measurements undertaken on the physical hardware, with no privileged vendor knowledge. In practice these sources play distinct roles. The documentation provides the architectural description and contract, namely instruction semantics, memory maps, soft-reset behaviour and unit descriptions. Wherever a behaviour is specified there we transcribe it directly, maintaining a provenance where numbers and semantics are taken from the documentation as written. Where the documentation is missing specific details, which is more the case for the Blackhole than the Wormhole, or ambiguous, we explored TT-Metal and Tenstorrent's Low Level Kernel (LLK) code bases. This was especially useful in debugging, for instance to understand how the packers and unpackers are driven and how the different configuration values actually map to application behaviour. Where it was not possible to determine behaviour based on these two sources we then resolved the question empirically on the hardware itself via bespoke probes.

Development of these hardware probes resulted in \textit{perfbench}, which is a suite of TT-Metal binaries that act as timing and investigation programs whose output is device cycle counts. These are constructed so that the identical binary runs both on silicon and the simulator, allowing the two to be compared to judge the simulator's accuracy. Three sub-suites comprise the benchmark:

\begin{itemize}
    \item Tensix unit costs providing per instruction metrics.
    \item Baby RISC-V front end including the issue rate, cost of pushing a Tensix instruction, branch and fetch behaviour, and the depth and sharing of the Tensix instruction queue.
    \item NoC behaviour which measures latency against hop count and against the number of links that concurrent flows share.
    \item DRAM rate behaviour, measuring bandwidth and latency of DRAM accesses.
\end{itemize}

Beyond calibrating the simulator, this benchmark has value in its own right because the resulting measurements are able to characterise aspects of the hardware which the documentation does not, and are the basis for several of the metrics reported in Section~\ref{sec:cycle_accurate}.

\subsection{Optimising Wolfpine's performance}
\label{sec:optimise_perf}
\label{sec:optimisation}
Wolfpine is written in pure Python, and this was a deliberate choice to lower the barrier to experimentation by the wider community. But the disadvantage is that of performance, especially when wishing to simulate the full device comprising many tiles. The naive execution model, in which every cycle ticks every active block in every tile, spends the overwhelming majority of its time ticking components whose state does not change. This is represented as \emph{Naive ticking} in Table \ref{tab:pref}, and we optimised this by reworking the clock design to be instead an event-driven clock pump which preserves the simple per-cycle behaviour of Section~\ref{sec:design} whilst eliminating this wasted work.

Our event-driven mechanism is opt-in, where each active block may report whether clock ticking it would be a no-op for observable state, and a per-tile clock stops ticking a tile once every component on it reports to be idle. Waking occurs by the arrival of a NoC message, a host access, or a reset transition and these are the only stimuli that can act upon a dormant tile. The runtime of this optimisation can be seen as \emph{tile idle gating} in Table \ref{tab:pref}, providing over a 55 times speed up. 

Once initialised, TT-Metal's firmware spins on a go-message from the host, and such a core is functionally idle yet retires instructions continuously. We therefore recognise the specific behaviour of this loop, which comprises repeated iterations of continuously executing loads and stores, and deactivate this until a command is sent from TT-Metal. The runtime based on this optimisation is reported as \emph{firmware-loop parking} in Table \ref{tab:pref}.

This was then refined to generalise the idle predicate into a statement of when a component next requires attention, such as a block occupied for a known number of cycles or a NoC packet with a known arrival time. This makes explicit the cycle at which a block will next becomes observable, and the pump strides directly to the earliest such deadline rather than stepping through cycles in which nothing can occur. This has a slight impact on performance, \emph{deadline striding} in Table \ref{tab:pref} but is a prerequisite for the cycle accuracy in Section \ref{sec:cycle_accurate} and diagnostics functionality in Section \ref{sec:observability}.
 
\begin{table}[htb]
\centering
\small
\begin{tabular}{@{}lrr@{}}
\toprule
\textbf{Configuration} & \textbf{Wall-clock (s)} & \textbf{Speedup} \\
\midrule
Naive ticking                    & 6586.7 & 1.0$\times$ \\
+ tile idle gating               & 118.5  & 55.6$\times$ \\
+ firmware-loop parking          & 13.6   & 484$\times$ \\
+ deadline striding              & 15.6   & 422$\times$ \\
\quad + cost model               & 21.2   & 311$\times$ \\
\bottomrule
\end{tabular}
\caption{Wolfpine optimisations on the \texttt{pad\_multi\_core}
benchmark simulating a p150 Blackhole, with four Tensix cores. Rows are cumulative.}
\label{tab:pref}
\end{table}

Lastly, enabling the cost model, \emph{cost model} in Table \ref{tab:pref}, also incurs some overhead but this is still significantly faster than the initial naive implementation. Crucially, these different configurations are opt-in on a component by component basis, where correctness does not depend on the optimisation and instead when adding a new block the decision can be made whether each mechanism is appropriate or not.

\subsection{Towards cycle accuracy}
\label{sec:cycle_accurate}
As described in Section~\ref{sec:design}, Wolfpine is cycle-driven and not cycle-accurate. In this approach ordering is well defined but the number of cycles consumed does not model hardware timing. However, our cycle-driven foundation enables the incremental adding of timing fidelity functionality, and as such we added an opt-in cost model to appropriate blocks in the simulator. This model is expressed as declarative tables of per-unit cycle costs, covering the Tensix back-end units on a per-instruction basis alongside the NoC, DRAM, the baby RISC-V cores and L1. A common loader turns this into occupancy where a unit charged $N$ cycles for an operation declares itself busy until that deadline and is not clock ticked before it, integrating directly with the event-driven pump described in Section \ref{sec:optimise_perf}. 

A cost model comprises not only a set of cycle counts, but furthermore semantics for each which describe the point in the pipeline at which those cycles are spent. Whilst it would be simplest to consume the cycles by occupying the issuing unit for the stated duration, we found this was overly naive as it frequently misattributed where the time occurs and consequently mispredicted overlapped execution. That could be a major limitation when leveraging the simulator to identify and diagnose runtime bottlenecks. We therefore spend each cost based on where its documented or measured stall location lies. The load latency of the baby RISC-V cores provides an example of this because charged as issue occupancy every load would stall the core for the full latency regardless of whether its result is required immediately. This would over-penalise code in which independent instructions have been scheduled after the load. We instead spend the stall cost as a load-use interlock, where subsequent load issues and execution continues, with the pipeline stalling only if those follow-on instructions read the destination register before the latency has elapsed. Consequently, a load whose result is not immediately consumed costs nothing beyond issue, matching the behaviour of the hardware. 

Our NoC model decomposes a packet's latency into a fixed endpoint cost, reflecting injection and ejection through the NIUs, plus a per-hop cost that is accrued as the packet visits each router on its journey. Hop counts are computed over the topology in a manner that matches the hardware, a wrapping torus in which each NoC routes in one direction only. Therefore a transfer that is issued in the \emph{wrong} direction travels the long way around the ring and is charged accordingly. Bandwidth is treated separately from latency and packets are decomposed into flits, the unit of flow control, and each flit occupies a link for one cycle. A large transfer therefore holds its path in proportion to its size, and when two transfers share a link the second observes the occupancy of the first. 

Every number used by Wolfpine is either based on that documented in \cite{ttisadoc} or derived from our own hardware measurements. The NoC and memory-path costs have been validated by our timing microbenchmarks, where the same binary is run on silicon and on the simulator and cycle counts then compared. The Tensix per-instruction costs are currently based on what has been documented in \cite{ttisadoc}, since validating them at that granularity would require detailed hardware traces which are not publicly available. Importantly, as our cost model is an add-on then the functional simulator itself is uneffected if this is disabled at runtime. 

Table~\ref{tab:perfbench} reports, for a set of operations, the hardware-measured average cycles per operation and the Wolfpine's predicted number of cycles based on our cost model. Also reported is the simulator's percentage error. All hardware runs are on a p150 Blackhole and is is measured over iterations of a loop of varying sizes. The factional numbers for hardware number are due to averaging in the presence of hardware jitter which the simulator does not experience. It can be observed that with the exception of integer divide, \emph{divu} RISC-V operation, our simulator's prediction is close to the actual hardware. This is because unlike the other operations, integer divide requires a varying number of cycles between 6 and 33 depending on the operands. Wolfpine models this range in its data but the cost model always assumes the low end.
 
\begin{table}[htb]
\centering
\small
\begin{tabular}{@{}lrrr@{}}
\toprule
\textbf{Operation} & \makecell{\textbf{Hardware} \\ \textbf{avg cycles}} & \makecell{\textbf{Simulator} \\ \textbf{cycles}} & \makecell{\textbf{Percentage}\\ \textbf{error}} \\
\midrule
FPU \texttt{MVMUL} & 1.061--1.067 & 1.00 & $6$ \\
THCON \texttt{ADDDMAREG} & 2.973 & 3.00 & $0.9$ \\
RISC-V multiply & 1.985 & 2.00 & $0.8$ \\
RISC-V L1 load & 8.098 & 8.00 & $1.2$ \\
RISC-V \texttt{divu} & 33.001 & 6.00 & $82$ \\
NoC latency/hop    & 8.85--9.03 & 9.00 & $0.3$ to $1.7$ \\
\bottomrule
\end{tabular}
\caption{Operation-level timing comparison against average actual cycle on p150 Blackhole and predicted cycles from Wolfpine's cost model.}
\label{tab:perfbench}
\end{table}

\subsection{Observability and diagnostics}
\label{sec:observability}
A simulator's advantage over silicon is that everything is visible and this is especially important for HPC developers who can potentially gain additional insights about bottlenecks in their code compared to running on the hardware. Consequently, we have focussed on developing tooling to exploit this visibility, ultimately positioning Wolfpine as a debugging and analysis instrument rather than merely for testing correctness. The foundation of this is a typed publish--subscribe event bus onto which architectural events are published and the events that we currently track are detailed in Table \ref{tab:events}.

\begin{table*}
\centering
\small
\begin{tabular}{@{}lp{3.1cm}p{3.4cm}p{9.0cm}@{}}
\toprule
\textbf{Event} & \textbf{Emitted by} & \textbf{Principal fields} & \textbf{Description} \\
\midrule
Instr & Baby RISC-V cores & PC, instruction, register writeback, stall cycles \& reason & Published at instruction retirement; register writeback enables differential debugging against a reference RISC-V simulator, and stall attribution is populated when the cost model is active. \\
Dispatch & Tensix frontend & Opcode, target unit, thread & A Tensix instruction leaving a per-thread frontend for a back-end unit. \\
Compute & Tensix back-end units & Opcode, unit, thread, duration & Execution of an operation on the FPU, SFPU, matrix, packer, unpacker, mover or THCON; duration is the unit occupancy under the cost model, zero denoting ``not modelled'' rather than one cycle. \\
Mem & Address spaces & Op, address, size, region, triggering PC & A read or write actioned by the lookup arbitrator, classified by target region (L1, DRAM, MMIO) and attributed where possible to the program counter of the issuing instruction. \\
NoC & NIUs & Phase, transaction type, source, destination, size, issue cycle & A NoC transaction observed at send and receive; the issue cycle allows flight time to be computed per packet. \\
Sync & Mailboxes, TTSync & Kind, detail & Synchronisation activity: mailbox sends and receives, TTSync signals and waits. \\
Lifecycle & Host interface, device control & Kind, detail & Coarse-grained progress: kernel launches and completions, reset transitions, host session activity. \\
Counter & Counter aggregator & Counter name, value, kernel & Periodic long-format samples of per-unit performance counters, one row per unit and counter at each flush boundary. \\
\bottomrule
\end{tabular}
\caption{Events published on Wolfpine's typed event bus. Every event carries the cycle of emission and a structured unit identifier (chip, core coordinates, unit). Writers subscribe to any subset by category.}
\label{tab:events}
\end{table*}

Our approach is that Wolfpine itself provides no viewers, instead targeting formats that existing tools ingest. To this end, data is outputted in a range of downstream tool formats, including line-delimited JSON for ad-hoc analysis, Perfetto-compatible timelines for visual inspection, Spike-compatible per-core commit logs enabling differential debugging against a RISC-V reference, columnar Parquet datasets of counters and NoC transactions for analytical queries, Cachegrind-format memory profiles, and LCOV source-level attribution mapping simulated execution back to kernel source lines. All these data writers subscribe independently to the event bus and may be enabled simultaneously without conflicting with each other. 

Interactive diagnostics complements the event streams where per-component tracing may be enabled selectively for each baby RISC-V core, each NoC, the Tensix co-processor and memory. This enables address ranges to be snooped to report accesses as they occur. Furthermore, a progress watchdog detects deadlock by periodically sampling architectural state, reporting a diagnosis when a kernel has hung. In addition to being useful for code developers, we have found these capabilities valuable during the development of the simulator itself, as divergence from hardware behaviour is most efficiently localised by inspecting these kinds of traces.

\subsection{Multi-architecture support}
Although development of Wolfpine initially targeted the Wormhole, the abstraction of Section~\ref{sec:design} was designed to generalise. This enabled support for the Blackhole to be added fairly easily using the same infrastructure. The simulator expresses architectures as profiles, with an example provided in Listing \ref{lst:blackhole_profile}. A profile is a collection of per-architecture hardware details including tile grid coordinates, memory map ranges, NoC parameters, ISA extensions and cost constants. Device construction is therefore generic and consumes the provided profile, with the device, NoC, RISC-V baby cores and Tensix co-processor all parameterised.

\begin{lstlisting}[language=Python, caption={The Blackhole architecture profile (abridged) providing per-architecture hardware facts, each transcribed from the open documentation or measurement.}, label={lst:blackhole_profile}]
BLACKHOLE_PROFILE = ArchProfile(
    name="blackhole",
    # NoC grid 17x12
    noc_grid_x=17, noc_grid_y=12,
    # Largest single NoC burst
    noc_max_burst_size=16384,
    # BH Tensix L1 SRAM: 1536 KiB
    tensix_l1_size=1536 * 1024,
    # Baby-core local memories, 
    # doubled vs Wormhole
    brisc_local_mem_size=8 * 1024,
    ncrisc_local_mem_size=8 * 1024,
    trisc_local_mem_size=4 * 1024,
    # Enlarged Tensix config state; Blackhole
    # register layout and STALLWAIT semantics
    tensix_cfg_state_size=56,
    tensix_thd_state_size=68,
    tensix_blackhole=True,
    # ISA extensions over RV32IM; guards make
    # unmodelled hardware (F, V) fail loudly
    baby_core_isa_extensions=("zba", "zbb",
        "zaamo", "zfh", "f_guard"),
    trisc2_isa_extensions=("v_guard",),
    # Worker-visible NoC coords of the 8 DRAM
    # channels, from the SoC descriptor
    dram_unified_coords=((0, 11), ...),
    dram_physical_noc1_coords=((0, 1), ...),
    # DRAM per channel: 4 GiB less the 16 MiB
    # register aperture
    dram_channel_size=0xFF00_0000,
)
\end{lstlisting}

Much of what we parameterise is based upon the differences between the Wormhole and Blackhole, ultimately to enable support for both in a common code base. An example is \emph{baby\_core\_isa\_extension} in Listing \ref{lst:blackhole_profile}, where the baby RISC-V cores in the Wormhole support only RV32IM, whereas on the Blackhole they support additional extensions. Within the simulator there is a simple RV32I core, providing an implementation of the base RISC-V ISA and then implementations for the different RISC-V extensions which can be provided to the constructor, with the op-decode functionality handing unmatched instructions to them for handling as appropriate. 

Anecdotally, the ease of which we were able to integrate the Blackhole demonstrates the generality of our block-based conceptual approach. Supporting a new architecture predominantly required transcribing hardware facts from the documentation into a profile, and performing measurements on hardware as appropriate.

\section{Evaluation}
\label{sec:evaluation}
 
All experiments use TT-Metal v0.74, Wolfpine release version 2.0, and Tenstorrent's \texttt{ttsim} release v1.9.6. Hardware results are measured on a Blackhole accelerator which reports 120 Tensix tiles organised as a 12$\times$10 logical worker grid, at a clock frequency of 1350\,MHz, with eight DRAM banks, firmware bundle 19.6.0 and KMD 2.10.0. Simulator results are measured on a 12th Gen Intel Core i7-12700H (16 logical CPUs) with 48\,GiB of RAM, and CPython 3.12.13. Identical binaries run on all three platforms, built once from an unmodified TT-Metal and the simulators are selected through environment variables, with slow dispatch mode used throughout.
 
Table~\ref{tab:benchmarks} lists the benchmarks used in this study which progressively add complexity and introduce further.
 
\begin{table}[htb]
\centering
\footnotesize            
\setlength{\tabcolsep}{3pt}   
\begin{tabular}{@{}lp{1.0cm}p{3.0cm}@{}}
\toprule
\textbf{Benchmark} & \textbf{Size} & \textbf{Subsystems introduced} \\
\midrule
\texttt{add\_2\_integers\_in\_riscv}& 1 uint32 & Baby cores, L1, launch/go path \\
\texttt{loopback} & 51000 bf16 & NIU programming, NoC, DRAM tiles \\
\texttt{eltwise\_binary} & 64 tiles, 32x32, bf16 & Unpack $\to$ FPU $\to$ pack pipeline, circular buffers \\
\texttt{matmul\_single\_core} & $128^3$ bf16 & Matrix unit, Dst accumulation \\
\texttt{matmul\_multi\_core} & $640^3$, bf16, 40 Tensix & Multi-tile dispatch, inter-tile NoC, interleaved DRAM \\
\bottomrule
\end{tabular}
\caption{Details of the benchmarks used for evaluation.}
\label{tab:benchmarks}
\end{table}

\begin{table}[htb]
\centering
\small
\begin{tabular}{@{}lccc@{}}
\toprule
& \multicolumn{3}{c}{\textbf{Result accuracy (PCC)}} \\
\textbf{Benchmark} & \textbf{Wolfpine} & \textbf{ttsim} & \textbf{p150}\\
\midrule
\texttt{add\_2\_integers}   & 1           & 1   & 1        \\
\texttt{loopback}           & 1           & 1    & 1       \\
\texttt{eltwise\_binary}    & 1           & 1    & 1       \\
\texttt{matmul\_single\_core}        & 0.9982 & 0.9981 & 0.9833 \\
\texttt{matmul\_multi\_core}  & 0.99997 & 0.99995 & 0.9821\\
\bottomrule
\end{tabular}
\caption{Result accuracy of both simulators and the p150 Blackhole against each program's own
host-computed analytical golden. Accuracy reported as Pearson Correlation Coefficient (PCC).}
\label{tab:simaccuracy}
\end{table}

Table \ref{tab:simaccuracy} reports accuracy of results for our benchmarks across the p150 Blackhole and both simulators. Result accuracy is measured by comparing against the program's host-computed \emph{golden} reference and reported as the Pearson Correlation Coefficient (PCC) where a value of 1 represent an exact match. It can be seen that both simulators and hardware deliver similar accuracy of results, and in-fact for matrix multiplication the host mismatch is not an error in the simulators or hardware, but instead as floating point arithmetic is not associative the order of operations matters and this is subtly different between the host, p150 and simulators.
 
Table~\ref{tab:sim_perf} reports wall-clock time per benchmark for both Wolfpine and Tenstorrent's \texttt{ttsim}. It can be observed that \texttt{ttsim} is faster throughout, by between roughly four and twenty times, and this is largely due to Wolfpine being written in Python rather than C++. We also report the slowdown of each simulator compared to running the same benchmark on the p150 Blackhole hardware, where it can be seen our simulator is between 3.5 and 192 times slower than the hardware. \texttt{ttsim} performs more favourably compared to the hardware, and indeed for \emph{eltwise\_binary} is actually faster. This is because with such a small problem size and relatively simple compute, the overhead of PCIe data transfer and kernel launching dominates. By contrast, much of this is avoided when running on a simulator. The widening performance gap for the \emph{matmul\_multi\_core} benchmark is not caused, as one might assume, by simulating multiple Tensix cores, but instead by a combination of a larger problem size and differences between how the simulators handle data movement. \texttt{ttsim} completes a NoC transaction immediately by copying the required data, whereas the cost tracking of Wolfpine means that as part of this moment it must also model packet flight, flit-level link occupancy and contention. With many cores concurrently streaming interleaved tiles from DRAM, the simulated data-movement RISC-V cores spend much of their time polling NIU completion counters, which must be simulated in Wolfpine cycle by cycle. Moreover, the deadline striding optimisation of Section~\ref{sec:optimisation} can only skip cycles when every tile is idle, which is rarely the case with many tiles.
 
\begin{table}[htb]
\centering
\small
\begin{tabular}{@{}lrrrr@{}}
\toprule
& \multicolumn{2}{c}{\textbf{Wall-clock (s)}} & \multicolumn{2}{c}{\textbf{Slowdown vs p150}} \\
\cmidrule(lr){2-3} \cmidrule(lr){4-5}
\textbf{Benchmark} & Wolfpine & ttsim & Wolfpine & ttsim \\
\midrule
\texttt{add\_2\_integers}     & 8.4 & 1.8  & 3.5$\times$ & 0.8$\times$ \\
\texttt{loopback}             & 8.8 & 2.4  & 3.7$\times$ & 1.0$\times$ \\
\texttt{eltwise\_binary}      & 14.2 & 2.4 & 3.8$\times$ & 0.6$\times$ \\
\texttt{matmul\_single\_core}   & 32.4      & 7.2  & 9.3$\times$ & 2.1$\times$ \\
\texttt{matmul\_multi\_core}  & 786       & 37.8 & 192$\times$ & 9.2$\times$ \\
\bottomrule
\end{tabular}
\caption{Performance of Wolfpine against Tenstorrent's ttsim and slowdown of both of these simulators compared to running on p150 Blackhole hardware.}
\label{tab:sim_perf}
\end{table}
 
Table~\ref{tab:obs_matrix} compares the ability to observe execution and introspect kernels across the hardware, Tenstorrent's \texttt{ttsim}, and Wolfpine. It can be observed that \texttt{ttsim} lacks many of these capabilities and indeed it is their deliberate policy to exposes no trace, dump or timeline entry point. Instead, their own documentation suggests that one can obtain this by modifying the source code. Tenstorrent provide a range of hardware performance and debugging tools, but it can be seen that these do not provide all the introspection capabilities that our simulator supports.
 
\begin{table}[htb]
\centering
\small
\begin{tabular}{@{}lccc@{}}
\toprule
\textbf{Capability} & \textbf{HW} & \textbf{ttsim} & \textbf{Wolfpine} \\
\midrule
Per-instruction retirement trace   & \xmark & \xmark & \cmark \\
Register writeback log (Spike diff)& \xmark & \xmark  & \cmark \\
NoC transaction stream             & P & \xmark & \cmark \\
Timeline export (Perfetto)         & \cmark & \xmark & \cmark \\
Source-level cycle attribution     & P & \xmark  & \cmark \\
Memory-range snooping              & \xmark & \xmark & \cmark \\
Deadlock diagnosis                 & \cmark & \xmark  & \cmark \\
Conformance/UB flagging            & \xmark & \cmark  & \cmark \\
\bottomrule
\end{tabular}
\caption{Introspection capability matrix. Where \cmark~means supported as a user-facing
facility, \xmark~is not available and P is partial.}
\label{tab:obs_matrix}
\end{table}

\subsection{Case study: Nekbone}

Nekbone is a mini-application derived from Nek5000, a Gordon Bell Prize-winning high-order incompressible Navier–Stokes solver that uses the spectral element method. It captures the core algorithmic structure shared by Nek5000 and many similar HPC codes, and solves a standard Poisson equation using the Conjugate Gradient (CG) iterative method with a simple preconditioner. Each iteration involves vector operations, matrix–matrix multiplications, and nearest-neighbour communication with the \emph{AX} kernel being the most computationally intensive aspect at 75\% of the runtime. This kernel evaluates the Poisson operator by performing a series of many small matrix multiplications per element, with each element's grid points determined by the cube of the polynomial order (16 in this work).

Work in \cite{arshad2026spectral} ported this \emph{AX} kernel to the Tenstorrent Wormhole, and we use this same kernel to explore the cost model approximation of Wolfpine against the p150 Blackhole. We therefore ported and ran the code on both the Blackhole and Wolfpine for four $16^3$ spectral elements, 16384 data points in total, over one Tensix tile. There are three phases to the kernel, and Table \ref{tab:nekbone} reports metrics for each phase and in total. There is around a 10\% difference between the p150 and Wolfpine's prediction for each phase, however taken as a whole it can be observed that the overall total prediction is within 4\% of the hardware.

\begin{table}[htb]
  \centering
  \small
  \begin{tabular}{@{}lrrrrr@{}}
  \toprule
  & \multicolumn{2}{c}{\textbf{Wolfpine}} & \multicolumn{2}{c}{\textbf{p150 Blackhole}} & \textbf{Error}\\
  \cmidrule(lr){2-3} \cmidrule(lr){4-5}
  \textbf{Pass} & \textbf{Cycles} & \textbf{Runtime} & \textbf{Cycles} & \textbf{Runtime} &  \% \\
                &                 & ($\mu$s)         &                 & ($\mu$s)         &               \\
  \midrule
  1   & 18{,}304 & 13.56 & 16{,}605 & 12.30 & $10.2$ \\
  2       & 11{,}926 &  8.83 & 13{,}120 &  9.72 & $9.1$  \\
  3    & 29{,}836 & 22.10 & 32{,}784 & 24.28 & $9.0$  \\
  \midrule
  \textbf{Total} & \textbf{60{,}066} & \textbf{44.49} & \textbf{62{,}509} & \textbf{46.30} & $\mathbf{3.9}$ \\
  \bottomrule
  \end{tabular}
  \caption{Nekbone (4 elements, $16^3$, single Tensix core) predicted cycles and runtime against that measured on p150 Blackhole hardware.}
  \label{tab:nekbone}
  \end{table}

\begin{figure*}[t]
  \centering
  \includegraphics[width=0.85\textwidth]{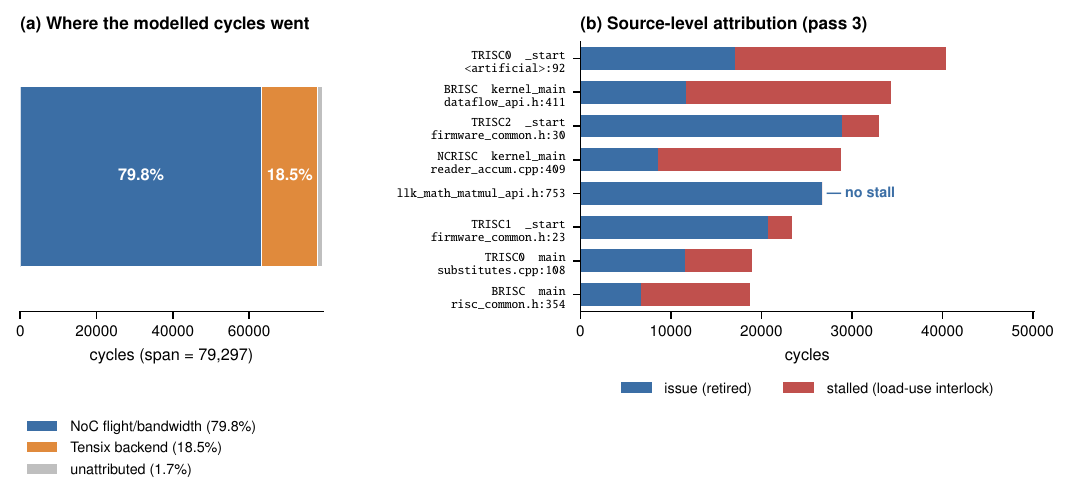}
  \caption{Cycle attribution for Nekbone from Wolfpine, \emph{(a)} is showing where cycles were spent between units and \emph{(b)} the heaviest source locations, split into cycles issuing and cycles stalled}
  \label{fig:nekbone-obs}
  \end{figure*}

Figure \ref{fig:nekbone-obs} illustrates a performance analysis report generated from Nekbone running via our Wolfpine simulator. In (a) this is showing what cycles were spent by what unit and (b) specific lines of code requiring the greatest runtime and how many cycles were active or stalled for each. 

\section{Conclusions and future work}

In this paper we have introduced Wolfpine, an open source simulator for AI accelerators and initially focussing on Tenstorrent's Tensix architecture. It supports both the Wormhole and Blackhole generations and integrates seamlessly with the TT-Metal software ecosystem. Constructed entirely from publicly available documentation, source code and hardware measurements, with no privileged vendor knowledge, Wolfpine enables code development without access to physical hardware and, through its opt-in cost model and observability tooling, provides introspection capabilities beyond those of either the silicon or the vendor's own simulator. The performance overhead of Python relative to Tenstorrent's C++ \texttt{ttsim}, even after the optimisation steps described in Section III-C, is a trade-off we accept in exchange for a modular, extensible codebase that is accessible to a wide community.

RISC-V is central to this work in two respects. Firstly, the openness of the standard makes it feasible to construct such a simulator without detailed proprietary knowledge of the microarchitecture, and Tenstorrent's own efforts to document their architecture and develop their software stack openly have been essential here. Secondly, because the base ISA is small and self-contained, with most functionality delivered through extensions, a single codebase can readily represent a range of hardware configurations, as demonstrated by the shared infrastructure supporting both the Wormhole and Blackhole through architecture profiles.

There are several directions for future work, the most immediate are refining the cost model where it diverges from silicon, such as the RISC-V integer division result of Section \ref{sec:cycle_accurate}, and modelling the ethernet tiles to support multi-chip configurations, which our single-chip evaluation does not yet exercise. Beyond Tenstorrent, our block-based abstraction was designed to generalise and we intend to target additional RISC-V based accelerators, including future product families such as Quasar as documentation emerges. A key question this will answer is how open an architecture must be, in terms of documentation, source code and measurable hardware, for a simulator of this kind to be constructed by outsiders. 

Finally, as a community developed tool we welcome external contributions, and hope the extensible design lowers the barrier to doing so. The simulator described in this paper is available open source at \url{https://github.com/hedgehope/wolfpine} and pip installed from the \emph{wolfpine} package on PyPI.

\section*{Acknowledgement}
This research was supported by an RSE personal research fellowship award number 3271 and the HPC-R project (EP/Z533701/1) . For the purpose of open access, the author has applied a Creative Commons Attribution (CC BY) licence to any Author Accepted Manuscript version arising from this submission.

\bibliographystyle{IEEEtran}
\bibliography{references.bib}

\end{document}